\documentclass{SciPost}

\hypersetup{
    colorlinks,
    linkcolor={red!50!black},
    citecolor={blue!50!black},
    urlcolor={blue!80!black}
}

\usepackage[bitstream-charter]{mathdesign}
\usepackage{hhline}
\usepackage{braket}
\usepackage{float}
\usepackage{comment}

\DeclareSymbolFont{usualmathcal}{OMS}{cmsy}{m}{n}
\DeclareSymbolFontAlphabet{\mathcal}{usualmathcal}

\fancypagestyle{SPstyle}{
\fancyhf{}
\lhead{\colorbox{scipostblue}{\bf \color{white} ~SciPost Physics }}
\rhead{{\bf \color{scipostdeepblue} ~Submission }}

\fancyfoot[C]{\textbf{\thepage}}
}

\newcommand{\da}{{{\downarrow}}}
\newcommand{\ua}{{{\uparrow}}}

\newcommand{\dl}{{{Z_L}}}
\newcommand{\dr}{{{Z_R}}}
\newcommand{\ep}{{{+}}}
\newcommand{\en}{{{-}}}
\newcommand{\Hloc}{\mathbf{h}}

\graphicspath{{./Figures/}}

\begin{document}

\pagestyle{SPstyle}

\begin{center}{\Large \textbf{\color{scipostdeepblue}{
Efficient search for excitable zero-modes in constrained systems\\
}}}\end{center}

\begin{center}\textbf{
Jean-Yves Desaules\textsuperscript{1$\star$}
}\end{center}

\begin{center}
{\bf 1} Institute of Science and Technology Austria (ISTA), Am Campus 1, 3400 Klosterneuburg, Austria
\\[\baselineskip]
$\star$ \href{mailto:email1}{\small jean-yves.desaules@ist.ac.at}
\end{center}

\section*{\color{scipostdeepblue}{Abstract}}
\textbf{\boldmath{%
    Kinetically constrained systems, such as those representing Rydberg atom arrays in the blockade regime, have gathered considerable attention due to the presence of atypical eigenstates in their spectrum. The latter manifests itself through the presence of quantum many-body scars as well as unusually large zero-mode (ZM) subspaces which contain analytically tractable eigenstates with various entanglement scalings. In particular, some of the latter are \emph{excitable} zero-modes (EZMs), meaning that they can be promoted to a non-zero energy for open boundary conditions. In this work, I present an efficient protocol for finding analytical expressions for translation-invariant EZMs in constrained systems, based on the eigendecomposition of the local unconstrained Hamiltonian. I  demonstrate the power of this method on a decorated Rydberg chain. In that model, my protocol directly produces a continuous matrix-product-state manifold located entirely in the zero-energy eigenspace for periodic boundary conditions. The span of that manifold grows exponentially with the system, and for odd system sizes it covers the \emph{entire} zero-momentum eigenspace with zero energy. I then show how the physical structure of the manifold, which is tied to the local eigenbasis, also allows one to derive analytical expressions for ZMs at non-zero momentum and for a polynomial number of exact scars at $E=\pm\sqrt{3}$. 
}}

\vspace{\baselineskip}

\vspace{10pt}
\noindent\rule{\textwidth}{1pt}
\tableofcontents
\noindent\rule{\textwidth}{1pt}
\vspace{10pt}


\section{Introduction}
The study of non-integrable quantum many-body systems at high temperature is notoriously difficult. While the eigenstate thermalisation hypothesis (ETH)~\cite{SrednickiETH,Deutsch2018ETH} allows to make some predictions about the expectation values of local observables for eigenstates, it also predicts that near the middle of the spectrum these states resemble random vectors in the Hilbert space. This lack of structure makes them analytically intractable beyond a few particles, and also makes their numerical study difficult. Indeed, the gap between them decreases exponentially fast, making techniques that are appropriate for ground states (e.g. the shift-inverse power method) unsuitable. Even in one dimension (1D), matrix-product-state (MPS) techniques quickly struggle as the entanglement entropy of these states obeys a volume law. This leaves exact diagonalisation (ED) as the main tool for their study, but this method too is badly affected by the exponential scaling of the Hilbert space.

As such, in the past decades there has been a growing interest in quantum many-body scars: non-thermal eigenstates embedded inside an otherwise thermalising spectrum~\cite{Serbyn2021Review,Chandran2023Review,Moudgalya2022Review}. These were first discovered in the PXP model~\cite{FendleySachdev,Lesanovsky2012}, which is an idealised representation of a 1D chain of Rydberg atoms~\cite{browaeys_many-body_2020,Bernien2017Rydberg,Bluvstein2021Controlling}. In that model, these atypical eigenstates are embedded across the energy spectrum and are responsible for coherent oscillations that were first witnessed in an experiment~\cite{Bernien2017Rydberg,TurnerNature,TurnerPRB}. While in the PXP case the scarred eigenstates are still not analytically tractable, in subsequent years many models were discovered that host ``exact'' scars, i.e. scars that admit an analytical description. In fact, additional scars were also discovered in the PXP model, but localised at zero energy~\cite{lin2019exact,mohapatra2025exact,ivanov2025exact,ivanov2025volume,karle2021arealaw,zhang2023extracting}. A major reason why these states stayed elusive for longer than their equivalents at finite energy density is because in the PXP model the zero-mode (ZM) subspace is exponentially degenerate. This can be proven analytically using the interplay between a chiral symmetry and the spatial symmetries of the system~\cite{schecter2018many,TurnerNature,TurnerPRB,buijsman2022number}. As a consequence, any numerical diagonalisation scheme will lead to a random mixture of states in the $E=0$ eigenspace. 

This exponential degeneracy is common to large families of models, and the search for analytically tractable ZMs has seen many developments over the recent years. This includes analytical construction based on the structure of Fock space itself~\cite{jonay2026cages,nicolau2026cages,tan2025cages,benami2025cages}, as well as numerical approaches that aim to extract simple ZMs from the degenerate eigenspace~\cite{karle2021arealaw,zhang2023extracting}.
Recently, a new telescopic sum framework has been proposed to show that a translation-invariant MPS is indeed an eigenstate of a local Hamiltonian for periodic boundary conditions (PBC)~\cite{ivanov2025exact,rubio2026local}. Crucially, under some conditions, this framework also allows to turn ZMs into eigenstates at non-zero energy for open boundary conditions (OBC). I will denote these special ZMs as \emph{excitable} zero-modes (EZMs).
This possibility of promoting the EZMs to finite energy makes them particularly interesting, as they can be used in conjunction with other ZMs to generate non-thermal dynamics that goes beyond the static behaviour of degenerate eigenstates. 

In Section~\ref{sec:search}, I will show that, in a family of constrained systems, the search for ZMs greatly simplifies in the case of EZMs due to their additional structure, which is unlocked by relying on the eigenbasis of the local unconstrained Hamiltonian. In fact, it allows to directly find analytical expressions for such eigenstates. In Section~\ref{sec:PXP}, I will then move on to the first example, which is the 1D PXP model. In that case, the proposed framework allows to effortlessly recover the unique known EZM. In Section~\ref{sec:PXP_star}, I will then move on to a similar model on a decorated lattice, and demonstrate that in that more complex case an \emph{exponentially} large family of EZMs can be unveiled. Finally, in Section~\ref{sec:db} I will show how the idea of relying on the structure of the unconstrained local Hamiltonian and only then applying the constraint can also be applied beyond translation-invariant MPSs to find exact eigenstates without a specific spatial structure. 

\section{A more efficient search for excitable zero-modes}\label{sec:search}
Before introducing the new EZM-search framework, I will start by briefly reviewing the main result of Ref.~\cite{ivanov2025exact} that I will rely on. It provides a criterion under which a translation-invariant MPS is an EZM of a constrained model based on a telescopic sum argument.   
\subsection{Review of the telescopic sum formalism}
Let me consider a one-dimensional Hilbert space where some local configurations in the computational basis are forbidden. Let me then consider a translation-invariant Hamiltonian that admits the form \begin{equation}\label{eq:H}
    H=\mathcal{P}\left(\sum_j \Hloc_j\right)\mathcal{P},
\end{equation}
where $\Hloc$ is the single-site Hamiltonian (that can be constraint violating) and $\mathcal{P}$ is the global projector on all allowed states~\footnote{While this equation seemingly gives rise to a non-local $H$ as  $\mathcal{P}$ acts on the whole chain, it can usually be written in a local fashion if $\mathcal{P}$ only enforces a local constraint.}. This formulation covers the PXP model and its extensions to longer range as well as to arbitrary spin-$S$. It also covers quasi-1D generalisations of it (e.g. ladders), provided that some sites are grouped together to recover a one-dimensional formulation. 

Let me denote the state of site $j$ in the computational basis by $s_j$ and the action of $\Hloc$ (ignoring the constraint for now) on it as $\Hloc\ket{s}=\sum_t \alpha_t^s \ket{t}$. One can now build a translation-invariant MPS where the state of each site is represented by the same tensor $M^s$, with the resulting wavefunction being 
\begin{equation}
    \ket{\psi_M}=\sum_{\{s_j\}}\mathrm{Tr}\left[M^{s_1}M^{s_2}\cdots M^{s_L}\right]\ket{s_1s_2\ldots s_L}
\end{equation}
for PBC.
If there exists an auxiliary matrix $X$ and a tensor $M^s$ such that 
\begin{align}
    &\left[X,M^s\right]=F^s=\sum_t \alpha_t^s M^t \ \forall s, \label{eq:C1} \\
    &M^{s_1}M^{s_2}\cdots M^{s_k}=0\  \text{if} \ s_1s_2\ldots s_k \ \text{is forbidden}, \label{eq:C2}
\end{align}
then the tensor $M^s$ is a translation-invariant EZM of $H$. This means that it is a zero-mode for PBC, while for OBC it allows to create eigenstates at finite energy. 
More precisely, if the auxiliary matrix $X$ has eigenvalues $\{\lambda_j\}$ with left/right eigenvectors $u_j^T/w_j$, then for any $l$ and $k$ there exists an MPS eigenstate at energy $\lambda_l-\lambda_k$. This eigenstate has the tensor $M^s$ in the bulk, with boundary vectors $u_l^T$ to the left and $w_k$ to the right~\cite{ivanov2025exact}. 

I will only give the main idea behind the proof, and readers are referred to Ref.~\cite{ivanov2025exact} for a formal demonstration. The key concept is that due to Eq.~\eqref{eq:C1}, acting with $\Hloc$ on site $j$ produces a term $\cdots\left(-M^sX\right)M^s\cdots$ while acting on site $j+1$ will instead produce $\cdots M^s\left(XM^s\right)\cdots $. As such, these two terms cancel. Globally applying $\sum \Hloc$ leads to a telescopic sum that annihilate the state, i.e. $\sum \Hloc\ket{\psi_M}=0$. Eq.~\eqref{eq:C2} then enforces that $\sum \Hloc\ket{\psi_M}$ indeed lives in the constrained space~\footnote{Note that the conditions on the commutation relations can be made less stringent if one only wants a ZM and not necessarily an EZM. However, as the focus is on the latter, only that case will be considered.}.

As noted in Ref.~\cite{ivanov2025exact}, finding such a set of $M^s$ and $X$ requires solving non-linear tensor equations and is generally NP-hard. As such, in that work an alternative method is proposed, relying instead on dimensionality reduction followed by alternating orthogonal projection. While that method is quite powerful and general, it suffers from two main drawbacks: the whole procedure is quite intricate, and the states produced at the end are still purely numerical, requiring them to be cast into an analytical translation-invariant MPS form ``by hand''. As such, if one only wishes to find the simplest states (which have a chance of being useful on an actual quantum device), this procedure is ill-suited. 

\subsection{EZM search in the local eigenbasis }
Instead, I propose the following procedure that allows to rapidly find exact translation-invariant EZMs by using the eigendecomposition of $\Hloc$. The basic principle behind this approach relies on the fact that Eq.~\eqref{eq:C1} requires the action of $[X,\cdot]$ on $M^s$ to be the same as the action of $\Hloc$ on $\ket{s}$. As such, the eigenvalues of both are necessarily related. Let $\epsilon_j$ and $v_j$ denote the eigenvalues and eigenvectors of $\Hloc$, one can then define a similar quantity 
\begin{equation}\label{eq:Vj}
    V^j=\sum_t(v_j)_tM^t, \quad \text{s.t.}\quad     [X,V^j]=\epsilon_j V^j.
\end{equation}
If I now denote by $\lambda_j$ the eigenvalues of $X$, it must hold that for any $\epsilon_j$ there exist two eigenvalues $\lambda_i$ and $\lambda_k$ for which $\lambda_i-\lambda_k=\epsilon_j$. As the eigenvalues and eigenvectors of $\Hloc$ are straightforward to find due to the small local dimension, this allows to essentially fix $X$ instead of having it as an unknown.

In this work, I will assume that the eigenvalues of $\Hloc$ are of the form $\{-E,0,+E \}$ (with some of them potentially degenerate), as this is the case for the vast majority of models fitting the framework of Eq.~\eqref{eq:H} with spin-1/2 and spin-1. I will then use the Ansatz that $X$ only has eigenvalues $E/2$ and $-E/2$ and has bond dimension $2m$.  Using the gauge freedom of the MPS, $X$ can then be fixed to take the form
\begin{equation}\label{eq:Xd}
    X=\eta
    \begin{pmatrix}
        I_m&0\\
        0&-I_m
    \end{pmatrix}.    
\end{equation}
with $\eta=E/2$~\footnote{As $X$ only appears in commutation relations, it can be shifted by any arbitrary scalar times the identity, and the choice to have it traceless is simply a convention.}. This guarantees that $X$ has the necessary eigenvalues to fulfil the desired relation with $V^j$. As the excited OBC energies of the EZM are also equal to differences of eigenvalues of $X$, this naturally implies that they must be equal to eigenenergies of $\Hloc$.

While in theory $X$ could have more sectors with a separation of $E$ between them, in practice this can be diagnosed rather easily. Indeed, if there are non-vanishing MPSs for which $r$ sectors (linearly separated by $E$) are necessary, then this directly implies the presence of eigenvalues $\pm (r-1)E$ in the OBC spectrum (see App.~\ref{app:r_sect}).  As the eigenvalues $\epsilon_j$ of $\Hloc$ are known, if one can diagonalise the system with OBC for a larger size, it is straightforward to check for which integer $r$ the spectrum contains the eigenvalues $\pm(r-1)\epsilon_j$. In the rest of this work, I will only discuss the case $r=2$ as shown in Eq.~\eqref{eq:Xd}. A brief discussion of $r>2$ can be found in Appendix~\ref{app:r_sect}.

Finally, coming back to Eq.~\eqref{eq:Xd}, while one could in theory take the two blocks to be of different size, the blocked structure of the $V^j$ that I will discuss below means that the imbalance between the blocks generally only adds nilpotent or inactive auxiliary directions that do not change the periodic MPS state.

While until now I have only used the relation between $\Hloc$ and $X$ to infer the structure of the latter, it also imposes a structure to the tensor $M^s$. Indeed, the diagonal structure of $X$ allows to conveniently represent each $M^j$ as split into four blocks of size $m$ by $m$, and one can use the relation of the $V^j$ defined in Eq.~\eqref{eq:Vj} to impose more structure. Denoting by $V^\pm$ the matrices corresponding to the eigenstates of $\Hloc$ with energy $\pm E$ and by $V^Z$ those corresponding to 0-energy eigenstates, it follows that 
\begin{equation}\label{eq:VjB}
    V^{+}=
    \begin{pmatrix}
        0&A\\
        0&0
    \end{pmatrix}, \quad
    V^{-}=\begin{pmatrix}
        0&0\\
        B&0
    \end{pmatrix}, \quad
    V^{Z}=
    \begin{pmatrix}
        C&0\\
        0&D
    \end{pmatrix}.
\end{equation}
The individual block matrices $A$ to $D$ are of size $m$ by $m$, and the $V^j$ (as well as the $M^s$ and $X$) have bond dimension $2m$. The form of Eq.~\eqref{eq:VjB} means that $V^+$ acts as a raising operator, taking the MPS from the $-E/2$ sector of $X$ to its $E/2$ sector, and that $V^-$ acts as a lowering operator. Meanwhile $V^Z$ only propagates the MPS in each sector without mixing them.

In case the eigenvalues have a higher multiplicity, there are then simply multiple tensors with the same block structure but different matrices inside the blocks. Finally, inverting the relation between the $M^s$ and $V^j$ gives the minimal structure required for the $M^s$ to satisfy the commutation relations of Eq.~\eqref{eq:C1}. The problem thus reduces to finding such block-matrices that satisfy the constraint in Eq.~\eqref{eq:C2}. This highlights the essence of this approach, which is to start from the space of zero-modes of the unconstrained Hamiltonian decomposed using the eigenbasis of $\Hloc$, and then look for states in that space that do satisfy the constraint.

The block form of Eq.~\eqref{eq:VjB} already has a few noteworthy implications. The main one is the nilpotency of $V^{\ep}$ and $V^{\en}$, as $V^{\ep}V^{\ep}=V^{\en}V^{\en}=0$. Actually, the condition is even stronger than this. Starting from $V^{\ep}$, as any $V^Z$ is purely block-diagonal, the next non-$V^Z$ block must necessarily be a $V^{\en}$. As these states are zero-modes, the number of $V^{\ep}$ and $V^{\en}$ must be the same, but this nilpotency indicates the stronger condition that they must necessarily alternate. This is a direct consequence of the choice of having $X$ with only two eigenvalue sectors, for a more detailed discussion see Appendix~\ref{app:r_sect}. 

\section{1D PXP model}\label{sec:PXP}
I will now demonstrate this approach on the PXP model, for which a single EZM is known, that is only invariant under translation by two sites.
\subsection{Model and properties}
The Hamiltonian is
\begin{equation}\label{eq:PXP}
    H_\mathrm{PXP}=\mathcal{P}_\mathrm{1D}\left(\sum_{j=0}^{L-1} \sigma^x_j\right)\mathcal{P}_\mathrm{1D}=\sum_{j=0}^{L-1} P_{j-1}\sigma^x_jP_{j+1}
\end{equation}
where the local projectors $P_j=\ket{\da}\bra{\da}_j$ enforce the Rydberg blockade that prevents neighbouring spins from being up at the same time, and where $\mathcal{P}_{1D}=\prod_{j=0}^{L-1}(1-\ket{\ua\ua}\bra{\ua\ua}_{j,j+1})$ is its global counterpart. Here, the single-site Hamiltonian $\Hloc$ is simply the $\sigma^x$ Pauli matrix. The projectors dressing it mean that it can flip any spin as long as the no-$\ua\ua$ constraint is not violated. The model in Eq.~\eqref{eq:PXP} is natively translation-invariant and fits into the form of Eq.~\eqref{eq:H}.
\subsection{Physical spin-1/2 sites}
I will start by considering the sites individually, without any grouping. The relevant local Hamiltonian is then simply $\Hloc=\sigma^x$. Its eigenvalues are then $\pm 1$ and its eigenstates imply
\begin{equation}
    V^{\ep}=\frac{M^{\da}+M^{\ua}}{\sqrt2},
    \qquad
    V^{\en}=\frac{M^{\da}-M^{\ua}}{\sqrt2},
\end{equation}
for the $V^j$ matrices.
This allows to define $X$ as in Eq.~\eqref{eq:Xd} with $\eta=1/2$ and -- using the structure of the $V^j$ in Eq.~\eqref{eq:VjB} and inverting the definitions for $V^{\ep}$ and $V^{\en}$ -- to write
\begin{equation}
    M^{\da}=
    \begin{pmatrix}
        0&\frac1{\sqrt2}A\\
        \frac1{\sqrt2}B&0
    \end{pmatrix}, \qquad
    M^{\ua}=
    \begin{pmatrix}
        0&\frac1{\sqrt2}A\\
        -\frac1{\sqrt2}B&0
    \end{pmatrix}.
\end{equation}
Finally, imposing the constraint
\begin{equation}
    M^{\ua} M^{\ua}
    =
    \begin{pmatrix}
        -\frac12 AB&0\\
        0&-\frac12 BA
    \end{pmatrix}=0
\end{equation}
implies $AB=BA=0$. However, this also forces
\begin{equation}
    M^{\da} M^{\da}
    =
    \frac12
    \begin{pmatrix}
        AB&0\\
        0&BA
    \end{pmatrix}
    =0,
\end{equation}
as well as $M^{\da}M^{\ua}=M^{\ua}M^{\da}=0$. As such, there is no EZM which is translation-invariant by one site, in line with the literature.

\subsection{Effective spin-1 sites}
However, one can extend the search by including MPSs which are only invariant under translation by two sites. To do this, consider the local coarse-grained basis which was previously introduced in Ref.~\cite{lin2019exact} where $\ket{O}=\ket{00}$, $\ket{L}=\ket{10}$ and $\ket{R}=\ket{01}$. The constraint now takes the form of forbidding $RL$ while $\Hloc$ becomes $\ket{O}\left(\bra{L}+\bra{R}\right)+\mathrm{h.c.}$ It is straightforward to diagonalise this operator, and using its eigenstates with eigenvalues ${-\sqrt{2},0,\sqrt{2}}$ one can set $\eta=1/\sqrt{2}$
and 
\begin{equation}
    \begin{aligned}
        V^{+}&{=}\frac{M^L{+}M^R{+}\sqrt2\,M^O}{2}, \\ V^{-}&{=}\frac{M^L{+}M^R{-}\sqrt2\,M^O}{2}, \\
        V^{Z}&{=}\frac{M^L{-}M^R}{\sqrt2}.
    \end{aligned}
\end{equation}
Using the block form of the $V^j$ in Eq.~\eqref{eq:VjB} and inverting the relation with the $M^s$ gives
\begin{equation}
        M^O{=}
        \begin{pmatrix}
            0&\frac{1}{\sqrt2}A\\
            -\frac{1}{\sqrt2}B&0
        \end{pmatrix}, \quad
        M^L{=}
        \begin{pmatrix}
            \frac{1}{\sqrt2}C&\frac12 A\\
            \frac12 B&\frac{1}{\sqrt2}D
        \end{pmatrix}, \quad
        M^R{=}
        \begin{pmatrix}
            -\frac{1}{\sqrt2}C&\frac12 A\\
            \frac12 B&-\frac{1}{\sqrt2}D
        \end{pmatrix}.
\end{equation}
Now the only step left is imposing the no-$RL$ condition $M^RM^L=0$.
As this matrix product leads to
\begin{equation}
    M^RM^L{=}\frac{1}{2}
    \begin{pmatrix}
        -C^2{+}\frac12 AB&
        -\frac{1}{\sqrt2}CA{+}\frac{1}{\sqrt2}AD\\[0.8em]
        \frac{1}{\sqrt2}BC{-}\frac{1}{\sqrt2}DB&
        \frac12 BA{-} D^2
    \end{pmatrix},
\end{equation}
setting it to zero is equivalent to satisfying the equations
\begin{equation}\label{eq:ABCD_1D}
    \begin{cases}
        AB&=2C^2, \\
        AD&=CA, \\
        BC&=DB, \\
        BA&=2D^2.
    \end{cases}
\end{equation}

Unlike in the single-site PXP case, this no longer has the trivial tensor $M^s=0$ as its only solution. However, satisfying all conditions in Eq.~\eqref{eq:ABCD_1D} directly implies $M^LM^R=0$ on top of $M^RM^L=0$. As such, EZMs which are translation-invariant by two sites can only occupy a subset of the Hilbert space which has this additional constraint. 

Despite Eq.~\eqref{eq:ABCD_1D} still being somewhat complex, one already gets a non-trivial solution for $m=1$ where the matrices $A$ to $D$ reduce to scalars (which I will denote by lowercase letters). Meanwhile, this is not possible in the original problem as formulated in Ref.~\cite{ivanov2025exact}, as setting the matrices $M^s$ and $X$ to scalars make all trivially vanish. In the case at hand, assuming $m=1$ allows to quickly assert that $c=d$, yielding the remaining equation $ab=2c^2$. Due to the gauge and normalisation freedom, any non-zero choice of these matrices leads to the same state. As such, one can pick an arbitrary solution, such as $a=b=\sqrt{2}$ and $c=d=1$. This gives 
\begin{equation}\label{eq:PXP_MPS}
    M^O=\begin{pmatrix}
        0&1\\
        -1&0
    \end{pmatrix}, \quad
    M^L=\frac{1}{\sqrt{2}}
    \begin{pmatrix}
        1&1\\
        1 &1
    \end{pmatrix}\!, \quad
    M^R=\frac{1}{\sqrt{2}}
    \begin{pmatrix}
        -1&1\\
        1&-1
    \end{pmatrix}.
\end{equation}
It exactly corresponds to the state $\ket{\Phi_1}$ of Refs.~\cite{lin2019exact,ivanov2025exact} given by
\begin{equation}\label{eq:Phi1}
    M_{\Phi_1}^O=\begin{pmatrix}
        0 & -1\\
        1 &0
    \end{pmatrix},\quad
    M_{\Phi_1}^L=\begin{pmatrix}
        0& 0\\
        0 &{-}\sqrt{2}
    \end{pmatrix},\quad
    M_{\Phi_1}^R=\begin{pmatrix}
        \sqrt{2}& 0\\
        0 & 0
    \end{pmatrix},
\end{equation}
after the transformation 
\begin{equation}
    M^s_{\Phi_1}=-GM^sG^{-1} \quad  \text{with} \quad
    G=\frac{1}{\sqrt{2}}\begin{pmatrix}
        1 & -1\\
        1 &1
    \end{pmatrix}.
\end{equation}
In fact, for any choice of matrices $(A,B,C,D)$ satisfying all relations in Eq.~\eqref{eq:ABCD_1D}, $M^s$ reduces to the tensor product of the unique bond-dimension 2 solution with an $m\times m$ matrix $K$ as $M^s_{m}= M^s\vert_{m=1}\otimes K$ (see  Appendix~\ref{app:1D} for the derivation). As such, $K$ is an irrelevant degree of freedom and there is only one solution. This matches with the known literature as well as diagonalisation of the system with OBC, which shows only one eigenstate at energy $\pm \sqrt{2}$. This case demonstrates how the procedure can allow to derive non-trivial solutions without having to actually solve any non-linear tensor equations, only scalar ones.

The eigenbasis procedure also gives a new perspective on the $\ket{\Phi_1}$ state. While the formulation of the latter provides a clearer picture in the computational basis, as it makes it obvious that no $RL$ nor $LR$ configurations occur, the solution as derived in Eq.~\eqref{eq:PXP_MPS} is instead simpler in the local energy basis. Indeed, it gives
\begin{equation}
    V^+=\begin{pmatrix}
        0&\sqrt{2}\\
        0&0
    \end{pmatrix}, \quad
    V^-=
    \begin{pmatrix}
        0&0\\
        \sqrt{2} &0
    \end{pmatrix}, \quad
    V^Z= 
    \begin{pmatrix}
        1 &0 \\
        0& 1
    \end{pmatrix}.
\end{equation}
Crucially, $V^Z$ is simply the identity matrix. This makes it clear that $V^\pm V^Z=V^\pm=V^Z V^\pm$. This implies that for any state with a $\ket{+Z}$ substring, there must be a state with the same background but with a $\ket{Z+}$ substring instead. While this relation can first seem fortuitous, it is actually a direct consequence of the constraint. Indeed, while $\ket{+Z}$ and $\ket{Z+}$ both violate the constraint, $\ket{+Z}+\ket{Z+}$ \emph{does not}, as the forbidden $\ket{RL}$ substring appears with prefactor $-1/\sqrt{8}$ in the former and $1/\sqrt{8}$ in the latter. Along with other such relations, this formulation highlights $\ket{\Phi_1}$ as the sole local superposition of ZM product states in the energy basis that lies within the no-$RL$ constrained space.

This also allows to understand why this state does not allow $LR$ on top of the the no-$RL$ constraint. Indeed, any two-site superposition of states with the same energy that cancels one will also cancel the other.

\subsection{Coarser graining}
Finally, I conclude this review of the 1D PXP case by discussing coarser graining of the original sites. One can for example wonder if grouping together $l$ sites leads to new EZMs that are only invariant under $T^{l}$. While finding exact solutions for such cases requires very careful bookkeeping due to the growing dimension of the local Hilbert space, ruling out their presence is straightforward. Indeed, diagonalising $\Hloc$ is still numerically easy, and one can then scan for the appearance of its eigenvalues in a larger system with OBC and $L=nl$ sites, with $n$ some integer. For example, if grouping together 4 sites, one finds that on top of $\{\pm \sqrt{2},0\}$, $\Hloc$ has the eigenvalues $\pm \sqrt{4\pm\sqrt{10}}$ which are not present for $l=2$. However, the spectrum of the PXP model with OBC (for $L=20$) does not reveal such new eigenvalues, only the ones at $\pm\sqrt{2}$ corresponding to the known states. The same is true for $l=3$ (and $L=18$). As such, the $l=2$ EZMs are likely the only ones in that model. 

\section{PXP model on a star lattice}~\label{sec:PXP_star}
Let me now move on to a case for which the presence of EZMs (and ZMs in general) has not been investigated before. I will show that it holds an exponential number of EZMs that admit a compact MPS description.
\subsection{Model and properties}
\begin{figure}[tb]
    \centering
    \includegraphics[width=0.8\columnwidth]{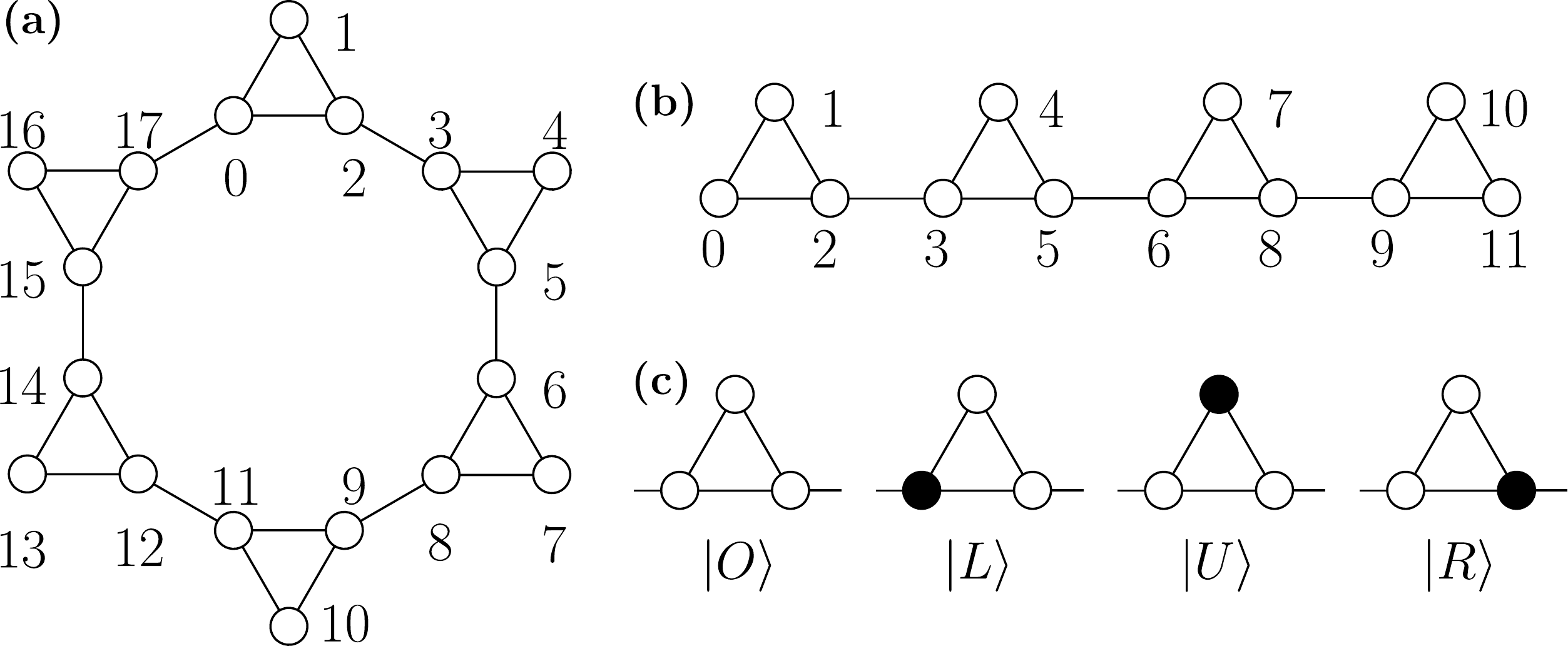}
    \caption{
        PXP model on a 1D star lattice. Down-spins and up-spins are denoted by ${\circ}$ and ${\bullet}$ respectively. (a) Chain with PBC, (b) chain with OBC, (c) triangle basis.}
    \label{fig:schematic}
\end{figure}

I  consider a decorated 1D lattice, where every other pair of sites has an extra site added to form a triangle, see Fig.~\ref{fig:schematic} (a) and (b) for the PBC and OBC geometries respectively. Let $L_\Delta=L/3$ denote the number of triangles, the Hamiltonian is then given by  
\begin{equation}\label{eq:star_PXP}
        H_{\star}{=}\sum_{j=0}^{L_\Delta-1}P_{3j-1}\sigma^x_{3j}P_{3j+1}P_{3j+2}+P_{3j}\sigma^x_{3j+1}P_{3j+2}
        +P_{3j}P_{3j+1}\sigma^x_{3j+2}P_{3j+3}
\end{equation}
for PBC, while for OBC the projectors $P_{-1}$ and $P_{3L_\Delta}$ are replaced by the identity.
This model is invariant under spatial reflection as well as under translation by 3 sites. It also has the chiral symmetry $\mathcal{C}=\prod_{j=0}^{3L_\Delta-1}\sigma^z_j$ which satisfies the anticommutation relation $\{\mathcal{C},H_\star\}=0$.
This model was already investigated in Ref.~\cite{verde2026engineering}, in which it was shown that scarring existed from the $\ket{\da\da\ldots \da}$ state, but also that it was non-integrable and had no symmetries beyond spatial ones, apart from the chiral spectral-reflection symmetry. 

As the Hamiltonian is only invariant under translation by 3 sites, in order to get a translation-invariant model I will group states in each triangle together, and denote their state using the local basis $\ket{O}$, $\ket{L}$, $\ket{U}$ and  $\ket{R}$ as shown in Fig.~\ref{fig:schematic} (c).
In that basis, the constraint only forbids the $RL$ configuration and $\Hloc=\ket{O}\left(\bra{L}+\bra{U}+\bra{R}\right)+\mathrm{h.c.}$ The Hamiltonian then becomes 
\begin{equation}\label{eq:star_PXP_tri}
    \begin{aligned}
        H_{\star}=\mathcal{P}_{\star}\left(\sum_{j=0}^{L_\Delta-1} \Hloc_j\right)\mathcal{P}_{\star}, \qquad \mathcal{P}_{\star}=\prod_{j=0}^{L_\Delta-1}\left(1{-}\ket{RL}\bra{RL}_{j,j+1}\right),
    \end{aligned}    
\end{equation}
thus fitting into the desired form of Eq.~\eqref{eq:H}.
\subsection{EZM search}
I will denote the states in the eigenbasis of $\Hloc$ as 
\begin{equation}\label{eq:db}
    \begin{aligned}
        \ket{\dl}&=\frac{\ket{L}-\ket{U}}{\sqrt{2}}, \quad
        \ket{\dr}=\frac{\ket{R}-\ket{U}}{\sqrt{2}} \\
        \ket{\ep}&=\frac{\ket{L}+\ket{U}+\ket{R}+\sqrt{3}\ket{O}} {\sqrt{6}} \\
        \ket{\en}&=\frac{\ket{L}+\ket{U}+\ket{R}-\sqrt{3}\ket{O}}{\sqrt{6}},
    \end{aligned}
\end{equation}
which have energies $0$, $0$, $\sqrt{3}$ and $-\sqrt{3}$. This basis is not orthogonal as $\braket{\dl|\dr}=1/2$, however it is still an eigenbasis of $\Hloc$ and minimises the constraint violation in the $E=0$ subspace as $\ket{\dl \dr}$, $\ket{\dl \dl}$ and $\ket{\dr \dr}$ are allowed. In fact, this eigenbasis already reveals two simple product ZMs
\begin{align}
    \ket{\mathcal{Z}_L}&=\ket{\dl\dl \cdots \dl}, \label{eq:ZZL}\\ \ket{\mathcal{Z}_R}&=\ket{\dr\dr \cdots \dr}. \label{eq:ZZR}        
\end{align}
Based on the previous decomposition, I set $\eta{=}\frac{\sqrt3}{2}$ and define $V^j$  with respect to the $M^s$ as in Eq.~\eqref{eq:db}. 
Due to the degeneracy of the 0 eigenspace, the $V^j$ now are
\begin{equation}
    \begin{aligned}
        V^{\ep}
        =
        \begin{pmatrix}
            0&A\\
            0&0
        \end{pmatrix},&
        \quad
        V^{\en}
        =
        \begin{pmatrix}
            0&0\\
            B&0
        \end{pmatrix}, \\
        V^{\dl}=
        \begin{pmatrix}
            C_{\dl}&0\\
            0&D_{\dl}
        \end{pmatrix},&
        \quad
        V^{\dr}=
        \begin{pmatrix}
            C_{\dr}&0\\
            0&D_{\dr}
        \end{pmatrix}.
    \end{aligned}
\end{equation}
Inverting the definition of the $V^j$ leads to 
\begin{equation}
    \begin{aligned}
        M^O&=\frac{1}{\sqrt{2}}
        \begin{pmatrix}
            0&A\\
            -B&0
        \end{pmatrix}, \\
        M^U&=\frac{1}{3\sqrt{2}}
        \begin{pmatrix}
            -2(C_{\dl}{+}C_{\dr})&\sqrt3A\\
            \sqrt3B&-2(D_{\dl}{+}D_{\dr})
        \end{pmatrix}, \\
        M^L&=\frac{1}{3\sqrt{2}}
        \begin{pmatrix}
            2(2C_{\dl}{-}C_{\dr})&\sqrt3A\\
            \sqrt3B&2(2D_{\dl}{-}D_{\dr})
        \end{pmatrix}, \\
        M^R&=\frac{1}{3\sqrt{2}}
        \begin{pmatrix}
            2(-C_{\dl}{+}2C_{\dr})&\sqrt3A\\
            \sqrt3B&2(2D_{\dr}{-}D_{\dl})
        \end{pmatrix}.
    \end{aligned}
\end{equation}
Finally, imposing the no-\(RL\) condition $M^RM^L=0$ yields
\begin{equation}
    \begin{cases}
        &4(-C_{\dl}+2C_{\dr})(2C_{\dl}-C_{\dr})+3AB=0,\\
        &(-C_{\dl}+2C_{\dr})A+A(2D_{\dl}-D_{\dr})=0, \\
        &B(2C_{\dl}-C_{\dr})+(-D_{\dl}+2D_{\dr})B=0, \\
        &3BA+4(-D_{\dl}+2D_{\dr})(2D_{\dl}-D_{\dr})=0.
    \end{cases}
\end{equation}

While one can once again find a non-trivial solution for scalar variables, I will directly provide the more general case and treat the scalar case later. For any $A$ and $B$, these equations are satisfied by 
\begin{equation}
    \begin{cases}
        C_{\dl}&=(A-2B)/\sqrt{12}, \\
        D_{\dl}&=(B-2A)/\sqrt{12}, \\
        C_{\dr}&=(2A-B)/\sqrt{12}, \\
        D_{\dr}&=(2B-A)/\sqrt{12}. \\
    \end{cases}
\end{equation}
The matrices in the energy basis are thus
\begin{equation}
    \begin{aligned}
        V^+=
        \begin{pmatrix}
            0&A\\
            0&0
        \end{pmatrix}, \quad
        V^{\dl}
        =
        \frac1{\sqrt{12}}
        \begin{pmatrix}
            A-2B&0\\
            0&B-2A
        \end{pmatrix}, \\
        V^-=
        \begin{pmatrix}
            0&0\\
            B&0
        \end{pmatrix}, \quad
        V^{\dr}
        =
        \frac1{\sqrt{12}}
        \begin{pmatrix}
            2A-B&0\\
            0&2B-A
        \end{pmatrix},
    \end{aligned}
\end{equation}
and in the physical basis they take the simple form
\begin{equation} \label{eq:star_MPS}
    \begin{aligned}
        M^O&=\frac1{\sqrt2}
        \begin{pmatrix}
            0&A\\
            -B&0
        \end{pmatrix}, \quad
        M^U=
        \frac1{\sqrt6}
        \begin{pmatrix}
            B{-}A&A\\
            B&A{-}B
        \end{pmatrix},  \\
        M^L&=
        \frac1{\sqrt6}
        \begin{pmatrix}
            -B&A\\
            B&-A
        \end{pmatrix}, \quad
        M^R=
        \frac1{\sqrt6}
        \begin{pmatrix}
            A&A\\
            B&B
        \end{pmatrix}.
    \end{aligned}
\end{equation}
The remaining virtual gauge and normalisation freedom preserving this form  means that any transformation $\{A\mapsto cS^{-1}AS, \ B\mapsto cS^{-1}BS\}$ for an invertible matrix $S$ and a scalar $c$ leads to the same state.

\subsection{MPS manifold properties}
If the matrices $A$ and $B$ are simultaneously diagonalisable, then the MPS leads to a state that is simply a sum of $m=1$ states. In that limiting case, the rescaling freedom can be used to set $b$ to 1, yielding the single-parameter family of states
\begin{equation}\label{eq:star_MPS_m1}
    \begin{aligned}
        M^O&=\frac1{\sqrt2}
        \begin{pmatrix}
            0&a\\
            -1&0
        \end{pmatrix}, \quad
        M^U=
        \frac1{\sqrt6}
        \begin{pmatrix}
            1{-}a&a\\
            1&a{-}1
        \end{pmatrix}  \\
        M^L&=
        \frac1{\sqrt6}
        \begin{pmatrix}
            -1&a\\
            1&-a
        \end{pmatrix}, \quad
        M^R=
        \frac1{\sqrt6}
        \begin{pmatrix}
            a&a\\
            1&1
        \end{pmatrix}.
    \end{aligned}
\end{equation}
Crucially, for any $a$ this MPS not only enforces the absence of $RL$ but also of $LR$. In fact, one can verify that taking $a\to 1/a$ (or exchanging $a$ and $b$ if $b$ is not set to 1) is equivalent to swapping L and R in the resulting state.
Numerical testing shows that the span of the $m=1$ family is equal to $L_\Delta$. In fact, the space of EZM with bond dimension equal or smaller than 2 can be expanded by adding either $\ket{\mathcal{Z}_L}$ or $\ket{\mathcal{Z}_R}$ as defined in Eqs.~\eqref{eq:ZZL} and Eqs.~\eqref{eq:ZZR}. While these two states are not inside the MPS family (or its span) for any $m$ , their superposition $\ket{\mathcal{Z}_L}+(-1)^{L_\Delta}\ket{\mathcal{Z}_R}$ is in the $m=1$ manifold in the limits $a\to \pm\infty$ and $a\to 0$. As such, the span of that manifold and of $\ket{\mathcal{Z}_L}$ has dimension $L_\Delta+1$ and spans the \emph{entirety} of the intersection between the zero-energy subspace and the no-$LR$ subspace.   

This special property of the $m=1$ cases allows to create a large family of perturbations that break various properties of the model without affecting the ZMs in it. Most notably, one can break the spatial symmetry as well as the chiral symmetry. Adding a perturbation of the form  \( \sum_{j=0}^{L_\Delta-1} \mu_j \ket{LR}\bra{LR}_{j,j+1} \) with $\mu_j\in[-W,W]$ breaks all of them, and also destroys the other ZMs. It thus results in a model that still has the same $L_\Delta+1$ translation-invariant ZMs despite having none of the structure usually expected to generate such states (see App.~\ref{app:pert} for details).

Interestingly, up to some gauge transformation and a rescaling of the tensors, this family of states encompasses a generalization $\ket{\Phi'_1}$ of the EZM $\ket{\Phi_1}$ found in Ref.~\cite{lin2019exact}. It occurs for 
$a=1/(2+\sqrt{3})$. Using the original gauge for $\ket{\Phi_1}$, the tensors $M$ for $\ket{O}$, $\ket{L}$ and $\ket{R}$ are the same as in the 1D PXP case in Eq.~\eqref{eq:Phi1} while 
\begin{equation} 
    M^U=\begin{pmatrix}
        0 & 1 \\ 1 & 0 \end{pmatrix}\quad \text{and} 
    \quad X=\begin{pmatrix}   -1/2 & 1/\sqrt{2} \\ 1/\sqrt{2} & 1/2 \end{pmatrix}.
\end{equation}
I note that to the best of my knowledge the other area-law zero-modes discussed in Ref.~\cite{ivanov2025exact} do not have an equivalent in the 1D star lattice, as discussed in Appendix~\ref{app:1D_gener}.

Beyond $m=1$, if $A$ and $B$ do not commute then there is generally no way to further reduce the MPS into a lower $m$ instances.
Taking into account the remaining gauge and normalisation freedom, the number of apparent continuous degrees of freedom is of the order of $m^2$. While determining the span of the manifold analytically for a given $m$ is hard, this can be achieved numerically very efficiently. With the addition of the $\ket{\mathcal{Z}_L}$ state, the span (across all $m$) has dimension 
\begin{equation}
    d^0_{\mathrm{TI}}(L_\Delta)=\frac{1}{L_\Delta}
    \sum_{j=1}^{L_\Delta}
    2^{\gcd(L_\Delta,j)},
\end{equation}
where $\gcd(x,y)$ denotes the greatest common divisor between the integer $x$ and $y$. For odd $L_\Delta$, this covers the \emph{entire}  translation-invariant zero-mode subspace, while for even $L_\Delta$ this still covers a finite fraction. A detailed breakdown of the span of the manifold for various $L_\Delta$ and $m$ can be found in Appendix~\ref{app:count}. 

\subsection{Dynamical signatures with OBC}~\label{sec:dyn}
\begin{figure}[tb]
    \centering
    \includegraphics[width=1\columnwidth]{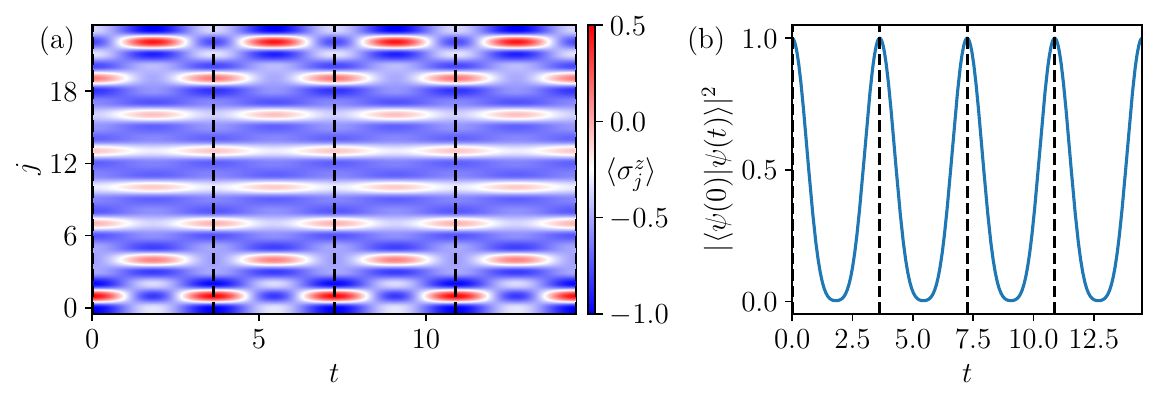}
    \caption{
        Dynamics in the 1D star lattice with 24 sites and OBC obtained through boundary excitations. The initial state has $m=2$ (bond dimension 4), both boundary vectors equal to $(1,1,1,1)/2$, and matrices $A=\begin{pmatrix} 0.244 &  0.402 \\
 0.0420 & 0.309\\ \end{pmatrix}$, $B=\begin{pmatrix}-0.163 & -0.227  \\ -0.411 & 0.374 
 \end{pmatrix}$. Both the local magnetisation (a) and the global return fidelity (b) exhibit the same periodic behaviour with period $T=2\pi/\sqrt{3}$ (vertical black dashed lines).}
    \label{fig:OBC_dyn}
\end{figure}

Due to the excitable nature of the ZMs in the manifold, for OBC it is trivial to construct eigenstates at energy $E=0$ and $E=\pm\sqrt{3}$. The only thing required is to choose boundary vectors that are eigenstates of $X$. As the latter has only two blocks of degenerate eigenvalues, as long as each boundary vector has non-zero entries in one block only, the resulting state will be an eigenstate.
Perhaps more interestingly, choosing both boundary vectors to have non-zero entries in both blocks will lead to a state having overlap on eigenstates at energies 0 and $\pm \sqrt{3}$. As such, it will oscillate indefinitely with a period $T=2\pi/\sqrt{3}$. An example of the resulting dynamics is shown on Fig.~\ref{fig:OBC_dyn} for $m=2$ with randomly sampled matrices $A$ and $B$.
While the bipartite von Neumann entanglement entropy is constant in time at a value of 0.7315, this does not mean that only spins belonging to the boundary triangles are unfrozen. Indeed, due to the correlated nature of the state the observables near the middle of the chain also oscillate as can be seen in panel (a), albeit with a reduced amplitude.
More details about the general structure of eigenstates with OBC can be found in Appendix~\ref{app:OBC}

\section{Exact eigenstates beyond translation-invariant MPS}\label{sec:db}
While the method described in Section~\ref{sec:search} only allows to efficiently find exact MPS expressions for translation-invariant EZM, the basic principle behind it can also be used to find non-translation-invariant eigenstates. 
Indeed, one can apply the same idea by creating superpositions of product states in the energy basis that do not violate the constraint.
For the 1D star lattice, recall the exact structure of the eigenstates of $\Hloc$ in Eq.~\eqref{eq:db}, which are $\ket{\ep}$, $\ket{\en}$, $\ket{\dl}$ and $\ket{\dr}$. I will call $\ket{\dl}$ and $\ket{\dr}$ dark states and $\ket{\ep}$ and $\ket{\en}$ bright states due to their energies.
Any string of these basis states is an eigenstate of the unconstrained problem with energy $E=\sqrt{3}\left(\# \ep-\# \en \right)$. The simplest states of that sort are $\ket{\mathcal{Z}_L}$ and $\ket{\mathcal{Z}_R}$ defined in Eqs.~\eqref{eq:ZZL} and \eqref{eq:ZZR}. Since they are both composed exclusively of dark states and do not violate the constraint, they are both product eigenstates of $H$ with 0 energy.

However, for other configurations of these basis states, things are not as simple. Unless consecutive sites are in a $\dl \dl$, $\dr \dr$ or $\dl \dr$ configuration, the constraint is violated. As seen in the 1D PXP case, this can be circumvented by making superpositions with another substring with the same energy but with a pattern that cancels out all the constraint-violating terms. For that, one can rely on the fact that the two superpositions
\begin{equation}\label{eq:db_sup}
    \ket{B_1 B_2}-\frac{1}{3}\ket{\dr\dl} \ \text{and} \ \ket{B_3 \dl}-\ket{\dr B_3}
\end{equation}
do not violate the constraint for any $B_i\in \{\ep,\en\}$. This can be checked easily from the definition of the energy basis states in Eq.~\eqref{eq:db} (and it trivially also holds for any state in the MPS manifold). 
While there are technically more superpositions that do not violate the constraint (e.g. $\ket{B_1 B_2}-\ket{\dr B_3}/\sqrt{3}$), these change the number of bright states by one. This means that the two states in the superposition do not have the same energy, and their superposition will not be an eigenstate. As such, I will only consider the superpositions in Eq.~\eqref{eq:db_sup} with $B_1\neq B_2$. This ensures that either the bright states stay the same or that a $\ep$ and an $\en$ are created/annihilated together, keeping $E=\sqrt{3}\left(\# \ep-\# \en \right)$ the same.

In order to find exact eigenstates, one can start from a product state in the energy basis and use the expressions in Eq.~\eqref{eq:db_sup} as replacement rules, i.e. 
\begin{align}
    &\ket{B_1 B_2} \rightarrow -\frac{1}{3}\ket{\dr\dl} \label{eq:rule1} \\ 
    &\ket{B_3 \dl} \leftrightarrow -\ket{\dr B_3},\label{eq:rule2}
\end{align}
which I will call contact moves and domain-wall moves respectively.
One can iteratively resolve constraint violations using these replacement rules. The eigenstate tied to that initial configuration is then the superposition of all states in the family generated with the correct prefactors. Notice that the first rule is non-reciprocal while the second one is. This is to avoid an undercounting of cases that would arise by connecting $\ket{+-}$ to $\ket{-+}$ through $-\frac{1}{3}\ket{\dr\dl}$ when it is not needed~\footnote{One can convince themselves of that by considering the case with $L_\Delta=2$, where $\ket{+-}-\frac{1}{3}\ket{\dr\dl}$ and $\ket{-+}-\frac{1}{3}\ket{\dr\dl}$ are both linearly independent eigenstates.}.

To have a valid zero-mode, it must hold that $\# \ep=\# \en$ but also that these states can only appear in alternating order. Indeed, any sequence containing $\ket{\ep D \cdots D \ep}$ with a dark state $D$ will be connected to a sequence with neighbouring $\ket{\ep \ep}$. To not violate the constraint, the latter will need a state with the same background but with  $-\frac{1}{3}\ket{\dr\dl}$ instead, which has a different energy. As such, any two $\ket{\ep}$ states must have a $\ket{\en}$ state between them. This alternating rule is actually the same one that naturally appeared by setting the auxiliary matrix $X$ with only two different eigenspaces in Eq.~\eqref{eq:Xd}.  

Finally, as I have made rule Eq.~\eqref{eq:rule1} unidirectional, only initial configurations without $\ket{\dr\dl}$ substrings are to be considered. One can then group all such valid starting configurations into equivalence classes, with two configurations being in the same class if they can be connected using solely domain-wall moves. As a convention, I will denote each class by the configuration where all dark states are $\ket{\dl}$ (with the exception of the $\ket{\mathcal{Z}_R}$ state which is its own representative)~\footnote{This is actually guaranteed in general as the only frozen $\dr$ are those with a $\ket{\dl}$ to their right which have been explicitly forbidden. Conversely, two valid states with only $\dl$ cannot be in the same equivalence class as any legal move would create a $\dr$.}. An example listing all equivalence classes for $L_\Delta=4$ is given in Appendix~\ref{app:count}

\begin{figure}[tb]
    \centering
    \includegraphics[width=\columnwidth]{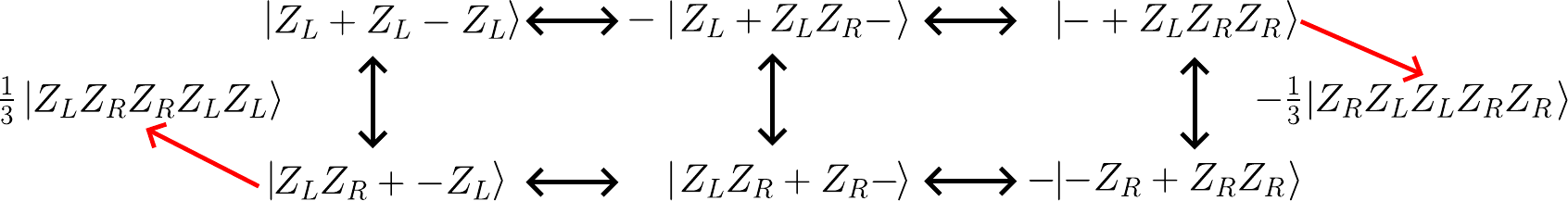}
    \caption{
        Structure of the zero-mode for the $\ket{\dl \ep \dl \en \dl}$ representative state. Black arrows denote domain-wall moves while red arrows denote contact moves. The 3 by 2 grid-like structure is due to the two bright sites being able to independently travel into the $\dl$ domains (of size 2 and 1) to their left using domain-wall moves. Once a bright site has reached the other through this, contact moves occur.}
    \label{fig:schematic_eig}
\end{figure}

Each equivalence class then leads to a linearly independent zero-mode that can be easily derived by tracking all the states connected by the rules in Eqs.~\eqref{eq:rule1} and \eqref{eq:rule2}, which also give the correct prefactor for each state. An example of such a derivation is shown in Fig.~\ref{fig:schematic_eig} for the $\ket{\dl \ep \dl \en \ep}$ representative state. The number of ZMs that can be obtained this way is then $d^0_{\mathrm{PBC}}(L_\Delta)=2^{L_\Delta}$, with $2\binom{L_\Delta}{2k}$ of them having a representative state with $2k$ bright states. 
This covers the full ZM space for $L_\Delta$ odd, while numerical results suggest that it still occupies a finite fraction $1/2<f<1$ of it in the even-$L_\Delta$ case (see App.~\ref{app:count}).

The number of translation invariant states that can be created out of these states can also be computed efficiently, this time using a mapping to binary necklaces. This gives 
\begin{equation}
d_{\mathrm{TI}}(L_\Delta)=
\frac{1}{L_\Delta}
\sum_{j=1}^{L_\Delta}
2^{\gcd(L_\Delta,j)},    
\end{equation}
matching the span of the ZM MPS manifold derived earlier (see App.~\ref{app:count} for details).

Overall, the ZMs derived in this fashion with zero momentum and those in the manifold span the same space. Both are obtained using the same idea of using the local unconstrained eigenbasis, but the results take a very different form. In fact going from one formulation to the other is a non-trivial task. Perhaps the best example of this is to be found in the exact PBC eigenstates that also exist at finite energy.

Indeed, the approach used in this last part can also be used to derive $L_\Delta$ independent PBC eigenstates at energies $\sqrt{3}$ and $-\sqrt{3}$. In order to see how that is possible, one needs to go back to the emergent conditions for bright states used for ZMs, which were that $\#\ep=\# \en$ and that the $\en$ and $\ep$ must alternate. While the former is only to ensure that the resulting state is a ZM, any configuration violating the latter cannot result in an eigenstate. As such, the only way to violate the first condition but not the second is to have a single $\ep$ or $\en$. The resulting eigenstates are of a simple form, however they \emph{cannot} be written down as an area-law translation-invariant MPS due to their nature resembling the $\ket{W}$-state. They instead require the twisted-translation-invariant framework of Ref.~\cite{ivanov2025exact} (see App.~\ref{app:finite_E} for details). This highlights that combining simple non-translation-invariant states into a translation-invariant MPS is not guaranteed to be a simple task. 

Finally, in the OBC case the same replacement-rule procedure used in this section can be applied. However, it requires re-deriving both the representative states and the states linked to them (see Appendix~\ref{app:OBC}). As such, the MPS formalism proves much simpler when going from PBC to OBC. It also provides a clear recipe for providing periodically oscillating states. 

\section{Conclusion}
Overall, in this work I have put forward a new framework for discovering area-law EZMs in a large class of constrained models based on the eigendecomposition of the local Hamiltonian. This method allowed me to quickly re-derive the only EZM in the 1D PXP chain, as well as to find a new exponentially large manifold of EZMs in the 1D star lattice. 

While the equations obtained through my framework are still non-trivial to solve, they have a simpler structure than the original formulation of Ref.~\cite{ivanov2025exact}. Crucially, the block parametrisation makes it possible to find non-trivial MPSs with a bond dimension equal to two by assuming that each block is a scalar. This makes the resolution of the equations well within the reach of automated symbolic solvers. For the more difficult case where the blocks are non-commuting matrices, at the time of writing state-of-the-art large language models are already capable of finding the solutions to the cases discussed in this work. As checking any solution can be done with very little overhead, I argue that solving such problems is an ideal use case for these programs.

I have also shown that the same idea of starting from eigenstates of the unconstrained problem and then finding solutions in the constrained subspace can be directly applied without using MPSs. The protocol then relies on finding local superpositions of states in the single-site energy basis that annihilate constraint-violating terms, and treating them as replacement rules. This approach allows to find non-translation-invariant ZMs, and generalises in a straightforward way beyond 1D and quasi-1D systems. However, in general it requires to find the relevant rules ``by hand'' for each model and leads to states that do not generalise as easily to OBC as the translation-invariant EZMs. As such, these two methods should be thought of as complementary rather than redundant. 

Finally, this work cements the PXP model on the 1D star lattice as a new playground for the study of ETH violation. Indeed, on top of the presence of an MPS manifold spanning exponentially many ZMs, it also holds a linear number of scarred eigenstates at energy $\pm \sqrt{3}$ for PBC. For OBC, thanks to the excitable nature of the ZMs, the subspaces at energies $\{-\sqrt{3},0,+\sqrt{3} \}$ all become exponentially large. 
While the dynamics that can be engineered using the EZMs has been briefly discussed in Section~\ref{sec:dyn}, there remains much to be studied. In particular, to what extent the dynamics can be tuned using the $A$ and $B$ matrices is a promising future direction of research. Another question is the relation to the EZMs discussed in this work to the scarring observed in Ref.~\cite{verde2026engineering}. 
It would also be interesting to further study the remaining ZMs which could not be explained through the local eigenstate approach for $L_\Delta$ even. While these are likely not excitable (see also Appendix~\ref{app:cg}), they could still have some tractable structure. For example, one of them is the volume-law $\ket{\Lambda}$ state of Ref.~\cite{ivanov2025volume}.

\section*{Acknowledgements}
I thank A. Kerschbaumer, E. Nicolau and M. Serbyn for useful discussions. \\
This work was funded in whole or in part by the Austrian Science Fund (FWF) 10.55776/COE1 and the European Union – NextGenerationEU. \\
Generative AI (OpenAI GPT-5.5) was used to help solve tensor equations and proofread the manuscript.

\begin{appendix}
\numberwithin{equation}{section}

\section{Larger number of sectors for X}\label{app:r_sect}
In this section, the implications of having an EZM for which $X$ needs more distinct eigenvalues are investigated. Importantly, I show how such a structure would imply specific eigenvalues for the same Hamiltonian with OBC.

Let me consider the case in which there are $r$ sectors with energy $E(\delta-k)$ with $k=0$ to $r-1$ with $\delta=(r-1)/2$. The matrix $X$ is then given by
\begin{equation}
    X
    =
    E
    \begin{pmatrix}
        \delta I_m&0&0&\cdots&0\\
        0&\left(\delta{-}1\right)I_m&0&\cdots&0\\
        0&0&\left(\delta{-}2\right)I_m&\cdots&0\\
        \vdots&\vdots&\vdots&\ddots&\vdots\\
        0&0&0&\cdots&-\delta I_m
    \end{pmatrix}.
\end{equation}
The presence of an EZM requiring such a larger number of sectors would imply the presence of eigenvalues $\pm qE$ in the OBC spectrum. Indeed, if that number of sectors is required, it means that there is a valid (i.e. with non-zero weight) sequence connecting sector 0 to $r-1$. By cutting the periodic trace at the endpoints of this segment and choosing boundary vectors in the eigenspaces \(k=0\) and \(k=r-1\), one obtains a non-zero OBC MPS. The usual commutator argument then gives
\begin{equation}
    H_{\rm OBC}\ket{\psi_M}=E[\delta{-}(\delta{-}r{+}1)]\ket{\psi_M}=E(r{-}1)\ket{\psi_M} .
\end{equation}
Therefore, if no such OBC eigenstate at energy \(E(r-1)\) exists, then no genuine PBC contribution requires all $r$ sectors and the $X$ tensor can be truncated.

Let me now study how the structure of the other matrices generalise for $r>2$.
As in the $r=2$ case, the desired relations are
\begin{equation}
    \begin{aligned}
        [X,V^{\ep}]&=E\,V^{\ep}, \\
        [X,V^{\en}]&=-E\,V^{\en}, \\
        [X,V^Z]&=0.
    \end{aligned}
\end{equation}
These relations force \(V^{\ep}\) to raise the \(X\)-charge by one step,
\(V^{\en}\) to lower it by one step, and \(V^Z\) to preserve it. Therefore
\begin{equation}
    \begin{aligned}
        V^{\ep}&=
        \begin{pmatrix}
            0&A_0&0&\cdots&0\\
            0&0&A_1&\cdots&0\\
            0&0&0&\ddots&0\\
            \vdots&\vdots&\vdots&\ddots&A_{r-2}\\
            0&0&0&\cdots&0
        \end{pmatrix}, \\
        V^{\en}&=
        \begin{pmatrix}
            0&0&0&\cdots&0\\
            B_0&0&0&\cdots&0\\
            0&B_1&0&\cdots&0\\
            \vdots&\vdots&\ddots&\ddots&0\\
            0&0&\cdots&B_{r-2}&0
        \end{pmatrix},\\
        V^Z&=
        \begin{pmatrix}
            C_0&0&0&\cdots&0\\
            0&C_1&0&\cdots&0\\
            0&0&C_2&\cdots&0\\
            \vdots&\vdots&\vdots&\ddots&0\\
            0&0&0&\cdots&C_{r-1}
        \end{pmatrix}.
    \end{aligned}
\end{equation}

The main consequence of this is that now the nilpotency of $V^{\ep}$ and $V^{\en}$ only forces $\left(V^{\ep}\right)^r=0$. As such, for any given $r$ the longest number of consecutive $V^{\ep}$ (with an arbitrary number of $V^Z$ between them) is $r-1$.

\section{Solution space for the 1D PXP model}\label{app:1D}
In this section, the full space of matrices that satisfy the desired commutation relations in the 1D PXP case is explored. Recall that the MPS takes the form
\begin{equation}
        M^O=
        \begin{pmatrix}
            0&\frac{1}{\sqrt2}A\\
            -\frac{1}{\sqrt2}B&0
        \end{pmatrix}, \quad
        M^L=
        \begin{pmatrix}
            \frac{1}{\sqrt2}C&\frac12 A\\
            \frac12 B&\frac{1}{\sqrt2}D
        \end{pmatrix}, \quad
        M^R=
        \begin{pmatrix}
            -\frac{1}{\sqrt2}C&\frac12 A\\
            \frac12 B&-\frac{1}{\sqrt2}D
        \end{pmatrix}.
\end{equation}
with the no-RL condition requiring
\begin{equation}
    AB=2C^2,
    AD=CA,
    BC=DB,
    BA=2D^2.
\end{equation}
The MPS gauge transformations preserving the \(X\)-block decomposition act as
\begin{equation}
    \begin{aligned}
        A\mapsto S^{-1}AT,
        \qquad
        B\mapsto T^{-1}BS, \\
        C\mapsto S^{-1}CS,
        \qquad
        D\mapsto T^{-1}DT.
    \end{aligned}
\end{equation}
Using that gauge freedom, one can set
\(
A=\sqrt2\,I
\) if $A$ is invertible, which is an assumption that can be made. Indeed, if $A=0$, this necessarily results in a state with norm zero. If instead $A$ has a non-zero nullspace, then by the same argument this nullspace cannot contribute to any valid MPS trajectory and can be truncated.    
Then \(
AD=CA
\)
implies
\(
D=C,
\)
and \(
AB=2C^2
\)
then gives
\(B=\sqrt2\,C^2.
\)
It is convenient to write
\(
C=D=K
\)
which leads to the solution \begin{equation}
    (A,B,C,D)
    =
    \left(
    \sqrt2\,I,\,
    \sqrt2\,K^2,\,
    \,K,\,
    \,K
    \right),
\end{equation}
for an arbitrary matrix $K$.
This leads to 
\begin{equation}
        M^O{=}
        \begin{pmatrix}
            0&I \\
            -K^2&0
        \end{pmatrix}, \quad
        M^L{=}\frac1{\sqrt2}
        \begin{pmatrix}
            K&I\\
            \,K^2&K
        \end{pmatrix}, \quad
        M^R{=}\frac1{\sqrt2}
        \begin{pmatrix}
            -K&I\\
            \,K^2&-K
        \end{pmatrix}.
\end{equation}
While this seemingly gives rise to a full family of eigenstates, the freedom of $K$ is actually trivial. Indeed, if $K$ is non-invertible then the space corresponding to the null eigenvalues can simply be truncated to get a new MPS with bond dimension 2$n$, with $n\leq m$.	Now assuming $K$ is invertible, the gauge transformation
\begin{equation} 
    \tilde{M}^s=G_KM^sG_K^{-1}\quad \text{with} \quad 
    G_K=\begin{pmatrix}
        K&0\\
        0& I
    \end{pmatrix}
\end{equation}
can be used to write each matrix as $\tilde{M}^s= M^s\vert_{m=1}\otimes K$, with the former being the bond-dimension 2 matrices given in Eq.~\eqref{eq:PXP_MPS}. Taking the trace of this object to compute the associated wave-function then leads to 
\begin{equation}
    \mathrm{Tr}\left[\prod_j\tilde{M}^{s_j}\right]=
\mathrm{Tr}\left[\prod_j M^{s_j}_{m=1}\right]\mathrm{Tr}\left[K^{L_\Delta}\right].
\end{equation}
As such, $K$ only contributes an irrelevant scalar prefactor and any $m>1$ reduces to the $m=1$ solution.

\section{Generalisation of 1D PXP eigenstates to the 1D star lattice}\label{app:1D_gener}
In the main text, I have discussed how the $\ket{\Phi_1}$ state can be promoted to an eigenstate $\ket{\Phi'_1}$ of the 1D star lattice by adding a suitable $M^U$ and changing the auxiliary matrix $X$. In fact, this can be done in two different ways. Taking $a=2+\sqrt{3}$ in Eq.~\eqref{eq:star_MPS_m1} gives a state $\ket{\Phi''_1}$ which has the same structure as $\ket{\Phi'_1}$ but with $L$ and $R$ swapped. While the same $LR$ swap can be done in the 1D PXP case, this is within the gauge freedom and leads to the same $\ket{\Phi_1}$ state. 
In both the 1D chain and the 1D star lattice, the basis states contributing to the original state and its $L\leftrightarrow R$ counterpart are the same, but their prefactors differ by a minus sign for each $\ket{O}$ present. It is straightforward to see that in the 1D PXP case, only states with an even number of $\ket{O}$ have non-zero contributions. However, in the 1D star lattice the number of $\ket{O}$ cells can take almost any value as they can be replaced by $\ket{U}$. As such, these two MPS correspond to the same state in 1D PXP, but in the 1D star lattice they are linearly independent. 

Beyond $\ket{\Phi_1}$, Ref.~\cite{ivanov2025exact} also discusses the states $\ket{\Phi_2}$, $\ket{\Theta_1}$ and $\ket{\Theta_2}$ as translation-invariant ZMs. As none of them are EZM, it is expected that if they have a generalisation in the 1D star lattice they would not be captured by the Ansatz used in this work. To check if such a generalisation exists, I have instead verified that there are no matrices $X$ and $M^U$ that satisfy the generalisation of Theorem 4 of that work in conjunction with the matrices $M^O$, $M^L$ and $M^R$ already given in Ref.~\cite{ivanov2025exact}. More precisely, there is no pair $X$ and $M^U$ for which
\begin{equation}
    \begin{aligned}
        [X,M^O]&=F^O \\
        M^L[X,M^L]&=M^LF^L \\
        M^O[X,M^L]&=M^OF^L \\
        M^U[X,M^L]&=M^UF^L \\
        [X,M^U]&=F^U \\
        [X,M^R]M^R&=F^RM^R \\
        [X,M^R]M^O&=F^RM^O \\
        [X,M^R]M^U&=F^RM^U,
    \end{aligned}
\end{equation}
with $F^O=M^L+M^U+M^R$ and $F^L=F^U=F^R=M^O$.
While this of course does not exclude a more complex generalisation, it does single out $\ket{\Phi_1}$ as the only area-law zero-mode that can be straightforwardly extended to the 1D star lattice by simply adding a suitable $M^U$.  I note that $\ket{\Phi_2}$ is a translation of $\ket{\Phi_1}$ by one spin-1/2 site in the 1D PXP case, and so by a half-triangle in the 1D star lattice. As the latter is not invariant under such a translation, it is not surprising that $\ket{\Phi_2}$ does not have an equivalent in this model.

Finally, beyond these area-law states, I note that the PXP model on the 1D star lattice also holds the volume-law state $\ket{\Lambda}$ of Ref.~\cite{ivanov2025volume} for $L_\Delta$ even.

\section{ZM-preserving perturbations}\label{app:pert}
In this section, I briefly discuss a few perturbations to the PXP model that preserve a subset of the zero-modes.
As stated in the main text, in the 1D star lattice the $m=1$ MPS manifold in Eq.~\eqref{eq:star_MPS_m1} has the special property of containing no $LR$ configuration on top of containing no $RL$ ones. As such, these states remain zero-modes of any Hamiltonian of the form
\begin{equation}
    H_\mathrm{LR}=H_{\star}+\sum_{j=0}^{L_\Delta-1} \mu_j P^{LR}_{j,j+1} h_j P^{LR}_{j,j+1},
\end{equation}
where $P^{LR}_{j,j+1}=\ket{LR}\bra{LR}_{j,j+1}$ is the projector on the LR substring on triangles $j$ and $j+1$, and $h_j$ is an arbitrary hermitian operator that can act on any site. This is reminiscent of the Shiraishi-Mori construction~\cite{Shiraishi17}, with the notable difference that $H_\star$ and $P^{LR}_{j,j+1}$ do not commute.

The simplest case is obtained by setting all $h_j$ to the identity, and the perturbation then reduces to \( \sum_{j=0}^{L_\Delta-1} \mu_j \ket{LR}\bra{LR}_{j,j+1} \). In both the homogeneous (all $\mu_j$ equal) and disordered case where each $\mu_j$ is uniformly sampled in $[-W,W]$, the full manifold (along with $\ket{\mathcal{Z}_{L/R}}$) are still exact zero-modes. This is despite the perturbation destroying the chiral symmetry, which leaves the $L_\Delta+1$ dimensional space as the sole ZMs. In the generic disordered case, all symmetries are actually destroyed and still, the $m=1$ manifold along with the all $\ket{\mathcal{Z}_{L/R}}$ states are exact zero-modes.

I note that adding such a perturbation in the 1D PXP case will then only stabilise the $\ket{\Phi_1}$ state while generically destroying all the other ZMs. One can also incorporate the no-$LR$ rule directly into the constraint. Consider the PXP model on a ladder, and group together each rung into the usual $L$, $R$ and $O$ states. While the constraint is that $LL$ and $RR$ are forbidden, if the ladder has an even number of rungs one can do the relabelling $L\leftrightarrow R$ on every other rung. The constraint is now that both $LR$ and $RL$ are forbidden. As such, the $\ket{\Phi_1}$ state will be an eigenstate at zero energy in the ladder, but the twist of half the sites means that it is now only invariant under translation by two rungs.

Finally, I note that in the 1D star lattice, the states $\ket{\mathcal{Z}_L}$ and $\ket{\mathcal{Z}_R}$ states are also eigenstates of the operators $N^\Delta_j=\sum_{k=3j}^{3j+2}n_k$ (with $n_k=\ket{\ua}\bra{\ua}_k$) which counts the number of up-spins on the $j$-th triangle. As such, adding a chemical potential $\lambda_j$ on each triangle to the Hamiltonian will still have these states as eigenstates, but at energy $\sum_{j=0}^{L_\Delta-1}\lambda_j$.

\section{ZM counting in the 1D star lattice}\label{app:count}
In this section, I discuss the number of ZMs in the 1D star lattice.
I first begin by showing an example of all the equivalence classes for $L_\Delta=4$ for the energy basis states discussed in Section~\ref{sec:db}. They are given below with the representative state in the first position, and ordered by the number $2k$ of bright states that this state has.

$k=0$
\begin{equation}
    \begin{aligned}
        \mathcal C_1&=\{\ket{\dl\dl\dl\dl}\},\\
        \mathcal C_2&=\{\ket{\dr\dr\dr\dr}\}.
    \end{aligned}
\end{equation}

$k=1$
\begin{equation}
    \begin{aligned}
        \mathcal C_3
        =&
        \{
        \ket{\ep\dl\dl\en},\,
        \ket{\dr\ep\dl\en},\,
        \ket{\dr\dr\ep\en}
        \},
        \\
        \mathcal C_4
        =&
        \{
        \ket{\dl\dl\en\ep},\,
        \ket{\ep\dl\en\dr},\,
        \ket{\dr\ep\en\dr}
        \},
        \\
        \mathcal C_5
        =&
        \{
        \ket{\ep\en\dl\dl},\,
        \ket{\ep\dr\en\dl},\,
        \ket{\ep\dr\dr\en}
        \},
        \\
        \mathcal C_6
        =&
        \{
        \ket{\dl\en\ep\dl},\,
        \ket{\dl\en\dr\ep},\,
        \ket{\ep\en\dr\dr}
        \},
        \\
        \mathcal C_7
        =&
        \{
        \ket{\dl\ep\en\dl},\,
        \ket{\dl\ep\dr\en},\,
        \ket{\en\ep\dr\dr}
        \},
        \\
        \mathcal C_8
        =&
        \{
        \ket{\dl\dl\ep\en},\,
        \ket{\en\dl\ep\dr},\,
        \ket{\dr\en\ep\dr}
        \},
        \\
        \mathcal C_9
        =&
        \{
        \ket{\en\dl\dl\ep},\,
        \ket{\dr\en\dl\ep},\,
        \ket{\dr\dr\en\ep}
        \},
        \\
        \mathcal C_{10}
        =&
        \{
        \ket{\en\ep\dl\dl},\,
        \ket{\en\dr\ep\dl},\,
        \ket{\en\dr\dr\ep}
        \}, \\
        \mathcal C_{11}
        =&
        \{
        \ket{\ep\dl\en\dl},\,
        \ket{\ep\dl\dr\en},\,
        \ket{\dr\ep\en\dl},\,
        \ket{\dr\ep\dr\en}
        \},
        \\
        \mathcal C_{12}
        =&
        \{
        \ket{\dl\en\dl\ep},\,
        \ket{\ep\en\dl\dr},\,
        \ket{\dl\dr\en\ep},\,
        \ket{\ep\dr\en\dr}
        \},
        \\
        \mathcal C_{13}
        =&
        \{
        \ket{\dl\ep\dl\en},\,
        \ket{\dl\dr\ep\en},\,
        \ket{\en\ep\dl\dr},\,
        \ket{\en\dr\ep\dr}
        \},
        \\
        \mathcal C_{14}
        =&
        \{
        \ket{\en\dl\ep\dl},\,
        \ket{\dr\en\ep\dl},\,
        \ket{\en\dl\dr\ep},\,
        \ket{\dr\en\dr\ep}
        \}.
    \end{aligned}
\end{equation}

$k=2$
\begin{equation}
    \begin{aligned}
        \mathcal C_{15}&=\{\ket{\ep\en\ep\en}\},\\
        \mathcal C_{16}&=\{\ket{\en\ep\en\ep}\}.
    \end{aligned}
\end{equation}
Given an initial state where the $\dl$ domains are of size $m_1$ to $m_l$, the equivalence class has $\prod_{j=1}^l (m_j+1)$ states as the bright term to the right of each domain can travel through it using domain-wall moves. 
The fact that (with the exception of the $\ket{\mathcal{Z}_R}$ state) in each class the representative state only has $\dl$ as dark states allows for an efficient counting of the number of equivalence classes, and thus of eigenmodes. For a state with $k$ $\ket{\ep}$ and $k$ $\ket{\en}$, there must be $2\binom{L_\Delta}{2k}$ valid representative states, with the binomial coefficient counting the number of ways of placing the $2k$ bright states and the factor of two accounting for the two ways ($\ep \en \ep \en$ and $\en \ep \en \ep$) of attributing $\ep$ and $\en$ to these states. For $k=0$, this instead accounts for the presence of the $\ket{\mathcal{Z}_R}$ state on top of the $\ket{\mathcal{Z}_L}$ one. Summing over all $k$ leads to a total of 
\begin{equation}
    d^0_{\mathrm{PBC}}(L_\Delta)=2\sum_{k=0}^{\lfloor L_\Delta/2 \rfloor }\binom{L_\Delta}{2k}=2^{L_\Delta}
\end{equation}
different equivalence classes, and thus ZMs. Comparing this value to the actual number of ZMs found numerically and given in Table~\ref{tab:ZM_num} shows that it covers the entire zero-mode subspace for $L_\Delta$ odd. In the even $L_\Delta$ case, the relation $2^{L_\Delta}<\#\mathrm{ZM}<2^{L_\Delta+1}$ holds for all accessible system sizes, suggesting that the states obtained using the replacement rules always account for a finite fraction of the ZMs. 
\begin{table}[H]
    \centering
    \begin{tabular}{c||c|c|c|c||c||c}
        $L_\Delta$ & $\{0,+\}$ & $\{0,-\}$ & $\{\pi,+\}$ & $\{\pi,-\}$ & Total & $d^0_{\mathrm{PBC}}$ \\
        \hhline{=#====#=#=}
        4 &  7 & 1 & 3 & 3 & 24 & 16\\
        5 & 1 & 7 & & & 32 & 32\\
        6 &  21 & 4 & 5 & 16 & 114 & 64 \\
        7 & 3 & 17 & & & 128& 128 \\
        8 & 65 & 19 & 49 & 33 & 480 & 256\\
        9 & 15 & 45 & & & 512 & 512 \\
    \end{tabular}
    \caption{Numerically computed number of ZMs in the different symmetry sectors $\{k,p\}$ of momentum and spatial inversion. The total value also includes the symmetry sectors $k\neq 0, \, \pi$ which are not shown individually.
    The last column shows the number $d^0_{\mathrm{PBC}}(L_\Delta)=2^{L_\Delta}$ of local energy basis ZMs, which account for all ZMs for $L_\Delta$ odd.}
    \label{tab:ZM_num}
\end{table}

The number of translation invariant states can then be computed using a mapping to binary necklaces. For each representative state, consider each $\ep$ to be a domain wall between a 0 and a 1 while each $\en$ is a domain wall between a 1 and a 0. The alternating nature of the bright states guarantees that this mapping works. Meanwhile, a $\dl$ implies no domain wall. For example, the representative state $\ep \dl \dl \en$ would correspond to $0111$ while $ \ep \dl \en \dl$ is equivalent to $0110$. The number of translation-invariant ZMs that can be built then given by the usual formula for binary necklaces as   
\begin{equation}
d^0_{\mathrm{TI}}(L_\Delta)=\frac{1}{L_\Delta}
    \sum_{j=1}^{L_\Delta}
    2^{\gcd(L_\Delta,j)}
\end{equation} 
For odd $L_\Delta$, this matches with the total number of ZMs found in the symmetry sectors with zero momentum (sum of the first two columns in Table~\ref{tab:ZM_num}). This can also be compared to the numerically-computed dimension of the span of the ZM manifold in Eq.~\eqref{eq:star_MPS}, which is shown in Table~\ref{tab:MPS_count}. The saturating dimension of the span is always one less than $d_{\mathrm{TI}}(L_\Delta)$. This is due to the $\ket{\mathcal{Z}_R}$ and $\ket{\mathcal{Z}_L}$ only being present in the superposition $(-1)^{L_\Delta}\ket{\mathcal{Z}_R}$ and not individually (or in the orthogonal superposition). Overall, one can conclude that both the MPS manifold and the translation-invariant superpositions of states in the energy basis span the entire translation-invariant ZM subspace for odd $L_\Delta$.

The results in Table~\ref{tab:MPS_count} also show that the value of $m$ at which the dimension of the span saturates is not fixed, but grows with $L_\Delta$. As such, as the system size gets larger, one can create EZMs that truly increase in complexity.
\begin{table}[H]
    \centering
    \begin{tabular}{c||c|c|c|c|c}
        $L_\Delta$ & $m=1$ & $m=2$ & $m=3$ & $m=4$ & $d^0_{\mathrm{TI}}$ \\
        \hhline{=#====#=}
        4 & 4 & \textbf{5} & 5 & 5 & 6 \\
        5 & 5 & \textbf{7} & 7 & 7 & 8\\
        6 & 6 & 12 & \textbf{13} & 13 & 14\\
        7 & 7 & 17 & \textbf{19} & 19 & 20 \\
        8 & 8 & 28 & \textbf{35} & 35 & 36\\
        9 & 9 & 39 & \textbf{59} & 59 & 60 \\
        10 & 10 & 60 & 106 & \textbf{107} & 108\\
        11 & 11 & 81 & 181 & \textbf{187} & 188\\
    \end{tabular}
    \caption{Dimension of the span of the zero-mode manifold in Eq.~\eqref{eq:star_MPS} with bond dimension $2m$ for various system sizes. The results in bold highlight the saturating $m$ for each $L_\Delta$, which is not fixed but grows with the system size.
        The last column indicates the number of translation-invariant eigenstates in the energy basis. The off-by-one result is due to the fact that $\ket{\mathcal{Z}_R}$ and $\ket{\mathcal{Z}_L}$ are not present independently in the manifold, but only as one linear combination.}
    \label{tab:MPS_count}
\end{table}

Finally, Table~\ref{tab:Chiral_count} shows the lower bound on the number of zero-modes due to the interplay between the chiral symmetry $\mathcal{C}=\prod_{j=0}^{3L_\Delta-1}\sigma^z_j$ and the spatial symmetries. As $\mathcal{C}$ simply gives the parity of the number of up-spins, the lower-bound is obtained as the absolute value of the difference between the number of states with an odd and an even number of such states in each symmetry sector. The symmetry sectors can be characterised by their momentum $k$ and, for $k=0$ or $\pi$, their eigenvalue $\pm 1$ under spatial inversion.
\begin{table}[H]
    \centering
    \begin{tabular}{c||c|c|c|c||c|}
        $L_\Delta$ & $\{0,+\}$ & $\{0,-\}$ & $\{\pi,+\}$ & $\{\pi,-\}$ & Total \\
        \hhline{=#====#=}
        4 &  5 & 1 & 3 & 1 & 16 \\
        5 & 1 & 1 & & & 2 \\
        6 &  13 & 2 & 3 & 12 & 62 \\
        7 & 1 & 1 & & & 2\\
        8 & 43 & 13 & 41 & 15 & 250 \\
        9 & 1 & 1 &  & & 2\\
        10 & 142 & 67 & 68 & 141 & 994 \\
    \end{tabular}
    \caption{Lower bound for the number of ZMs due to the interplay of spatial and chiral symmetries in the different symmetry sectors $\{k,p\}$ of momentum and spatial inversion. The total value also includes the symmetry sectors $k\neq 0, \, \pi$ which are not shown individually.}
    \label{tab:Chiral_count}
\end{table}
Overall, the actual number of ZMs (as given in Table~\ref{tab:ZM_num}) greatly exceeds the lower bound, both for even and odd $L_\Delta$. The fact that for odd system sizes this lower bound is constant is in stark contrast with the 1D PXP model, where it grows exponentially~\cite{TurnerPRB,buijsman2022number}.

\section{Eigenstates at non-zero-energy in the 1D star lattice with PBC}\label{app:finite_E} 
By considering states with an unequal number of $\ep$ and $\en$, one can theoretically create eigenstates with any energy that is a multiple of $\sqrt{3}$. However, as discussed in the main text, having two $\ep$ without a $\en$ in between (or the reverse) means that all states in the family will not have the same energy. As such, the only way to construct eigenstates with a finite energy using the method in Section~\ref{sec:db} is to have either a single $\ep$ or a single $\en$. Such states can be built using the same rules as the ZMs from the representatives states with a single $\pm$ in a background of $\dl$. There are then
\begin{equation}
    d^\pm_{\mathrm{PBC}}=\binom{L_\Delta}{1}=L_\Delta
\end{equation}
such representative states with energy $\sqrt{3}$ and the same number with $-\sqrt{3}$. The state corresponding to the representative with $\pm$ at site 0 (and $\dl$ everywhere else) can then be written as
\begin{equation}
    \ket{E^{\pm}_0}=\sum_{j=0}^{L_\Delta-1}\left(-\ket{\dr}\right)^{\otimes j} \ket{\pm} \ket{\dl}^{\otimes L_\Delta-1-j}.
\end{equation}
This in turns allows to write the state for which the representative has the $\pm$ at site $j$ as 
\begin{equation}
    \ket{E^{\pm}_j}=T_\Delta^{j}\ket{E^{\pm}_0},
\end{equation}
with $T_\Delta=T^3$ the translation by one triangle. 
It is also possible to create states with a desired momentum $k$ as
\begin{equation}
    \ket{\mathcal{E}^{\pm}_k}=\frac{1}{\sqrt{L_\Delta}}\!\!\sum_{j=0}^{L_\Delta-1}e^{-ikj}T_\Delta^{j}\ket{E^{\pm}_0}=\frac{1}{\sqrt{L_\Delta}}\!\!\sum_{j=0}^{L_\Delta-1}e^{-ikj}\ket{E^{\pm}_j}.
\end{equation}
Despite being translation-invariant, the $\ket{\mathcal{E}^{\pm}_{k=0}}$ states cannot be written as a translation-invariant MPS with size-independent bond dimension due to their $\ket{W}$-state nature where the sole $\pm$ state is fully delocalised. However, it can be written as a twisted-translation-invariant (TTI) MPS using the framework of Ref.~\cite{ivanov2025exact}. 

It is best defined directly in the energy basis, where
\begin{equation}
    \begin{aligned}
        V^{\dl}=\begin{pmatrix}
            1 & 1 \\ 0 & 0 
        \end{pmatrix}, \ 
        V^{\dr}=\begin{pmatrix}
            0 & 0 \\ 0 & -1 
        \end{pmatrix}, \ V^{+}=V^{-}=0, \
        V^{\dl}_1=V^{\dr}_1=V^{-}_1=0, \
        V_1^{+}=\begin{pmatrix}
            0 & 0 \\ 1 & 1 
        \end{pmatrix},
    \end{aligned}
\end{equation}
and with the twist matrix 
 \begin{equation}
    \mathcal{T}=\begin{pmatrix}
    0 & \mathbf{1}_{2 {\times} 2} \\ 0 & 0 \end{pmatrix}.
\end{equation}
The equivalent for the eigenstate with $E=-\sqrt{3}$ simply switches $V_1^{+}$ and $V_1^{-}$. 
From there one can also express the state directly into the computational basis, where the state with energy $\pm \sqrt{3}$ can be written as
\begin{equation}
    \begin{aligned}
        M^{L}=\frac{1}{\sqrt{2}}\begin{pmatrix}
            1 & 1 \\ 0 & 0 
        \end{pmatrix}, \ 
        M^{R}=\frac{-1}{\sqrt{2}}\begin{pmatrix}
            0 & 0 \\ 0 & 1 
        \end{pmatrix}, \\
        M^O=0, \ M^{U}=\frac{1}{\sqrt{2}}\begin{pmatrix}
            -1 & -1 \\ 0 & 1 
        \end{pmatrix} \\
        M^{L}_1=M^{R}_1=M^{U}_1=\frac{1}{\sqrt{6}}\begin{pmatrix}
            0 & 0 \\ 1 & 1 
        \end{pmatrix}, \\
        M_1^{O}=\frac{\pm1}{\sqrt{2}}\begin{pmatrix}
            0 & 0 \\ 1 & 1 
        \end{pmatrix}.
    \end{aligned}
\end{equation}

\section{Exact eigenstates in the 1D star lattice with OBC}\label{app:OBC}
In this section, I focus on the PXP model on a 1D star lattice with open boundary conditions.

First of all I discuss the span of the translation-invariant MPS manifold in Eq.~\eqref{eq:star_MPS}. For a single $(A,B)$ instance, there are $2m^2$ independent choices of boundary vectors that lead to a zero-mode. Indeed, there are two choices of sectors (the number of distinct eigenvalues of $X$), and in each sector there are $m$ choices for the right eigenvector and $m$ choices for the left one. As $X$ is diagonal, one can simply chose the unit vectors as boundary vectors. This counting matches with the numerics until the saturation value of the span is reached. The value of that span as well as the bound dimension 
$m^{1}_\mathrm{sat}$ at which it occurs (which simply corresponds to $\left\lceil \sqrt{\mathrm{Span}/2}\right\rceil$) is shown in Table~\ref{tab:ZM_MPS_OBC}. 
Interestingly, now computing the span over many generic $(A,B)$ instances does not lead to a larger saturation value, only to a much smaller value of $m$ required, which is denoted by $m^{\infty}_\mathrm{sat}$ in Table.~\ref{tab:ZM_MPS_OBC}

By choosing the left eigenvector in the $\pm \sqrt{3}/2$ sector and the right one in the $\mp \sqrt{3}/2$ sectors, one can also create exact eigenstates with $E=\pm \sqrt{3}$. For a given $(A,B)$ instance, the boundary vector dimension is now $m^2$, which again matches with numerical results up to saturation to $2^{L_\Delta-1}$. As for the ZMs, expanding to additional $(A,B)$ instance does not lead to a larger span but to a lower required $m$. The $m$ saturation values in both cases are the same as for the ZMs and are shown in Table.~\ref{tab:ZM_MPS_OBC}

\begin{table}[H]
    \centering
    \begin{tabular}{c||c|c||c|c}
        $L_\Delta$ & $0$ Span & $\pm$ Span & $m^{1}_\mathrm{sat}$& $m^{\infty}_\mathrm{sat}$ \\
        \hhline{=#==#==}
        4 &  27 & 15 & 4 & 2 \\
        5 & 58 & 31 & 6 & 3 \\
        6 &  121 & 63 & 8 & 3  \\
        7 & 248 & 127 & 12 & 3\\
        8 & 503 & 255 & 16 & 3 \\
    \end{tabular}
    \caption{Numerically computed span of the translation-invariant MPS manifold with boundary vectors for OBC, with energies 0 and $\pm \sqrt{3}$, and  value of $m$ needed to reach with either a single generic $(A,B)$ instance ($m^{1}_\mathrm{sat}$) or arbitrarily many $(A,B)$ instances ($m^{\infty}_\mathrm{sat}$).}
    \label{tab:ZM_MPS_OBC}
\end{table}

Importantly, checking the dimension of the span in the $E=\pm\sqrt{3}$ case against ED data shows that the span actually occupies the \emph{entire} eigenspace. The ED data as well as its split between the two symmetry sectors of spatial reflection is shown on Table.~\ref{tab:ZM_num_OBC}. That table also shows the dimensions  $d^{0}_\mathrm{OBC}$ and $d^{\pm}_\mathrm{OBC}$, which are respectively the number of eigenstates at energy $0$ and $\pm \sqrt{3}$ obtained using the replacement rules in the energy basis (see Section~\ref{app:db_OBC} below). Crucially, the span of these states has the same dimension as that of the MPS manifold in both cases, highlighting that they are different formulations of the same construction.
\begin{table}[H]
    \centering
    \begin{tabular}{c||c|c|c||c||c|c|c||c}
        $L_\Delta$ & $0,+$ & $0,-$ & $0$ & $d^{0}_\mathrm{OBC}$ & $\sqrt{3},+$ & $\sqrt{3},-$  & $\sqrt{3}$ & $d^{\pm}_\mathrm{OBC}$ \\
        \hhline{=#===#=#===#=}
        4 &  16 & 15 & 31 & 27 & 9 & 6 & 15 & 15 \\
        5 & 33 & 33 & 66 & 58 & 19 & 12 & 31 & 31\\
        6 &  78 & 77 & 155 & 121 & 28 & 35 & 63 & 63\\
        7 & 154 & 154 & 308 & 248 &71 & 56& 127 & 127\\
        8 & 343 & 342 & 685 & 503 & 120 & 135 & 255 & 255\\
    \end{tabular}
    \caption{Numerically computed number of states at $E=0$ and at $E=\sqrt{3}$ in the symmetry sectors $+1$ and $-1$ of spatial inversion and in total for the 1D star lattice with OBC. The number of states in the dark/bright basis at $E=0$ ($d^0_\mathrm{OBC}$) and at $E=\pm\sqrt{3}$ ($d^\pm_\mathrm{OBC}$) as computed in Section~\ref{app:db_OBC} is also given. While the latter matches with actual number of eigenstates at that energy, the former falls short by an exponential number of states.}
    \label{tab:ZM_num_OBC}
\end{table}

Finally, for completeness the lower bound of the number of ZMs due to the interplay of the chiral symmetry $\mathcal{C}$ and spatial inversion is shown in Table~\ref{tab:ZM_chiral_OBC}. As for PBC, the actual number of ZMs is much larger than this bound.
\begin{table}[H]
    \centering
    \begin{tabular}{c||c|c||c}
        $L_\Delta$ & $+$ & $-$ & Total  \\
        \hhline{=#==#=}
        4 &  8 & 3 & 11 \\
        5 & 3 & 3 & 6 \\
        6 & 24 & 17 & 41 \\
        7 & 4 & 4 & 8  \\
        8 & 81 & 72 & 153 \\
        9 & 5 & 5 & 10 \\
        10 & 291 & 280 & 571
    \end{tabular}
    \caption{Lower bound on the number of ZMs from the interplay of chiral symmetry and spatial inversion with OBC in both symmetry sectors of spatial inversion.}
    \label{tab:ZM_chiral_OBC}
\end{table}

\subsection{Replacement rule equivalence classes for OBC}\label{app:db_OBC}
As for ZMs with PBC and $L_\Delta$ odd and for PBC eigenstates with $E=\pm \sqrt{3}$, for OBC the full eigenspace with $E=\sqrt{3}$ can be described in the energy basis using replacement rules.
The rules in Eqs.~\eqref{eq:rule1} and \eqref{eq:rule2} to cancel out the blockade violation remain the same, and so does the need to have the $\ep$ and $\en$ alternate. As such, the only change in the problem is to be found in the equivalence classes and their representatives. 
Indeed, it is no longer possible to get rid of all $\dr$ cells through domain-wall moves, and they may accumulate at the right boundary. However, as all of them can be pushed there, each equivalence class has a representative consisting of a sequence of length $n$ over $\{\dl,\ep,\en\}$ followed by a string of $\dr$ of length $L_\Delta-n$. As such, accounting for all valid sequences for $n=0$ to $L_\Delta$ gives the full set of equivalence classes.

For eigenstates with energy \(+\sqrt3\), there must always be one more $\ep$ than $\en$. This means that in the no-$\dr$ sequence the bright terms must be $\ep \en \ep \cdots \en \ep$ (and the same with $\ep$ and $\en$ swapped for $-\sqrt{3}$). As this also implies an odd number of bright terms, for any given $n>0$ (the energy imbalance is impossible with $n=0$) there are $2^{n-1}$ different sequences. Indeed, the first $n-1$ sites can be either $\dl$ or a bright site, while the parity of the number of bright sites determines what the last site is. As such, for a chain of length $L_\Delta$ there are 
\begin{equation}
    d^{\pm}_\mathrm{OBC} (L_\Delta)=\sum_{n=1}^{L_\Delta}2^{n-1}=2^{L_\Delta}-1.
\end{equation}
This matches with the numerically computed number of eigenstates at that energy as shown in Table.~\ref{tab:ZM_num_OBC}.

A similar counting can be done for zero-modes, however it falls short of the numerically computed total (as for the PBC with even $L_\Delta$). One can use the same arguments as before, but this time the number of $\ep$ and $\en$ must be equal.
As before, for any $n>1$ there are $2^{n-1}$ ways of arranging the symbols. There is now also a contribution of 1 from $n=0$. Finally, for all states except one (the $\dl$-only state) in sectors with $n>0$ there are two possible bright symbol sequences, leading to $2^n-1$ possibilities. As such, the total number of equivalence classes is 
\begin{equation}
    d^0_\mathrm{OBC} (L_\Delta)=1+\sum_{n=1}^{L_\Delta}\left(2^{n}-1\right)=2^{L_\Delta+1}-\left(L_\Delta+1\right).
\end{equation}

\section{Coarser graining in the 1D star lattice}\label{app:cg}
In the rest of this work I have used triangles as effective sites in order to have a translation-invariant Hamiltonian. However, it is possible that some EZMs that are only invariant under translation by two triangles exist. One can gain insight on their existence by numerically diagonalising the Hamiltonian for $L_\Delta=2$ and checking if these eigenvalues appear in the spectrum of larger system with OBC. While this is indeed the case for some of the eigenvalues, they appear for OBC system size with $L_\Delta$ both odd and even. As such, their presence is likely only due to the dressing of the $L_\Delta=2$ system with some ``inert'' ZM states. The simplest cases would be states like $\ket{\dl}^{\otimes l} \ket{E_2}_{l,l+1}\ket{\dr}^{\otimes{L_\Delta-2-l}}$, where $\ket{E_2}_{l,l+1}$ denotes the energy $E_2$ eigenstates of the $L_\Delta=2$ system. In fact, this kind of construction guarantees that if the system with $L_\Delta$ and OBC has an eigenvalue $E$, this eigenvalue must appear with multiplicity at least 2 in the system for $L_\Delta+1$ and OBC.

\end{appendix}





\bibliography{biblio.bib}


\end{document}